%% file: main.tex
\documentclass[sigconf, screen]{acmart}
\AtBeginDocument{%
  }

\setcopyright{acmcopyright}
\acmConference{...}
\acmBooktitle{...}
\acmPrice{...}
\acmDOI{...}
\acmISBN{...}

\usepackage{tabularx}
\usepackage{makecell}

\renewcommand\footnotetextcopyrightpermission[1]{}

\begin{document}

\title{Leveraging Code-Mixed Product Metadata and User Feedback for Personalized Recommendation on Daraz Bangladesh}

\author{KM Fahim A Bari}
\affiliation{%
 \institution{East West University}
 \city{Dhaka}
 \country{Bangladesh}}
 \email{kmfahim2002@gmail.com}

 \author{Muhammad Abdullah Adnan}
\affiliation{%
  \institution{Bangladesh University of Engineering and Technology}
  \city{Dhaka}
  \country{Bangladesh}}
\email{adnan@cse.buet.ac.bd}

\author{Nafis Sadeq}
\affiliation{%
  \institution{East West Univeristy}
  \city{Dhaka}
  \country{Bangladesh}}
\email{nafis.sadeq@ewubd.edu}

\renewcommand{\shortauthors}{Bari et al.}

\begin{abstract}
Bangladeshi e-commerce platforms host millions of product reviews written in Bengali Unicode, English, and Banglish, where Bengali is phonetically transcribed in Latin script. However, the impact of code-mixed reviews on recommendation performance remains largely unexplored. We present the first such benchmarking on product reviews from Daraz Bangladesh, evaluating six model families under a per-user chronological leave-last-out protocol. To address the severe long-tail sparsity of the dataset, where 59.3\% of users have exactly one interaction, we conduct a systematic k-core threshold ablation across five density configurations. The results reveal that Item-based Collaborative Filtering remains stable across settings, Implicit Matrix Factorization degrades sharply with decreasing density, and Explicit Matrix Factorization uniquely improves at higher thresholds. To characterize the impact of code-mixing on recommendation quality, we perform a language-stratified evaluation of content-based filtering using character n-gram TF-IDF profiles. The results provide empirical evidence that fragmentation of the Banglish vocabulary reduces NDCG@10 by 46.8\% relative to Bengali-script users, a degradation traceable to transliteration inconsistency across surface forms. This work establishes a reproducible evaluation foundation for recommendation research in code-mixed, low-resource e-commerce settings. The code is publicly available at \url{https://github.com/os-car-war-thy/daraz-recsys}.
\end{abstract}


\maketitle

\input{Sources/introduction}

\input{Sources/dataset}
\input{Sources/experiments}
\input{Sources/related}
\input{Sources/conclusion}

\bibliographystyle{ACM-Reference-Format}
\bibliography{references.bib}

\end{document}

%% file: Sources/introduction.tex
\section{Introduction}

Recommender systems underpin product discovery in modern e-commerce~\cite{ricci2010introduction}. Existing canonical testbeds---MovieLens \cite{10.1145/2827872} and Amazon Product Reviews \cite{10.1145/2872427.2883037}---have driven substantial progress in recommendation research. However, these benchmarks are predominantly derived from mature Western digital marketplaces, leaving rapidly growing e-commerce ecosystems in emerging markets comparatively underrepresented.

Bangladesh presents a compelling case. With a population exceeding 170 million people \cite{united_nations_world_2024} and a rapidly expanding digital commerce sector, the country represents one of the largest emerging consumer markets in South Asia, with its e-commerce market projected to reach approximately USD 10 billion~\cite{bangladeshi-ecommerce}. Yet, little is known about how recommendation systems perform in such environments, where consumer behavior, product ecosystems, and user-generated content differ substantially from those represented in existing benchmarks. In particular, online reviews in Bangladesh frequently exhibit a distinctive code-mixing phenomenon \cite{shamael2024banglishrevlargescalebanglaenglishcodemixed,10129187}, where users seamlessly combine Bengali and English and often write Bengali words using Latin characters (commonly known as Banglish). For example, ``valo'', ``bhalo'', ``balo'', and ``vhalo'' are common Banglish variants of the Bengali word meaning ``good'' in English. These variants appear 19,834 times collectively in BanglishRev, a product review dataset from Daraz \cite{shamael2024banglishrevlargescalebanglaenglishcodemixed}. Such linguistic variation is not merely incidental noise; it introduces vocabulary fragmentation and representation challenges that are rarely encountered in conventional recommendation benchmarks and can substantially affect the modeling of textual side information \cite{10129187,ISLAM2024100069}.

To date, no systematic recommendation benchmark exists for Bengali e-commerce product data . Existing Bangladeshi e-commerce datasets have been developed primarily for sentiment analysis \cite{shamael2024banglishrevlargescalebanglaenglishcodemixed,10456378,9392733,9033741} and lack the user--item interaction structure required for recommendation research. BanglishRev dataset~\cite{shamael2024banglishrevlargescalebanglaenglishcodemixed} contains user--item interaction signal but the authors focused solely on sentiment analysis and did not explore product recommendation.

Our work makes three contributions:
\begin{itemize}
\item \textbf{First systematic benchmark.} We evaluate six model families on BanglishRev, establishing the first reproducible Top-N recommendation benchmark for Daraz product recommendation.
\item \textbf{Code-mix impact analysis.} A language-stratified evaluation of content-based filtering provides the first empirical measurement of how Banglish vocabulary fragmentation degrades recommendation quality, with Banglish-dominant users experiencing a 46.8\% reduction in NDCG@10 relative to Bengali-script users.
\item \textbf{Sparsity ablation.} We systematically vary k-core filtering thresholds across five density configurations, revealing that Implicit Matrix Factorization (ImplicitMF) degrades sharply with decreasing density, whereas Explicit Matrix Factorization (ExplicitMF) uniquely improves. This finding has important implications for model selection in sparse interaction settings.

\end{itemize}

%% file: Sources/dataset.tex
\section{Dataset and Preprocessing}

\subsection{BanglishRev}

BanglishRev \cite{shamael2024banglishrevlargescalebanglaenglishcodemixed} contains 3.24 million ratings from 934,949 users across 128,494 products collected from Bangladeshi online retail platforms. After deduplication---retaining the highest observed rating for each $(\text{user}, \text{item})$ pair as the strongest preference signal---the interaction table contains 2.67 million unique interactions. The resulting user--item matrix has a density of 0.0022\%, highlighting the extreme sparsity of the dataset.

User activity exhibits a pronounced long-tail distribution: 59.3\% of users have exactly one interaction and 75.9\% have at most two interactions, while only 1.5\% have more than 20 interactions. Item popularity is similarly skewed: 35.4\% of products receive a single review and 50\% receive at most two reviews. The rating distribution is strongly positive-skewed, with 77.93\% of ratings assigned the maximum five-star score, consistent with the voluntary review behavior observed on other e-commerce platforms. Furthermore, 45.89\% of the 3.24 million reviews contain no review text, creating a substantial challenge for text-based recommendation approaches.

\textbf{Language and script characteristics.}
To obtain reliable estimates of user language behavior, script composition is analyzed on the primary $(10,3)$ k-core filtered dataset, where each user has at least 10 interactions and each item has at least 3 interactions after convergent filtering. A user's dominant script is defined as the script accounting for the largest proportion of their reviews. Among the 39,550 retained users, 61.1\% wrote no review text (dominant script: empty), 23.0\% are English-dominant, 13.5\% Bengali-dominant, and 2.2\% Banglish-dominant. Only 27.6\% of users are effectively monolingual (more than 90\% of reviews in a single script), confirming the prevalence of multilingual and code-mixed behavior within the platform.

\subsection{Preprocessing}

\textbf{K-Core Filtering:}
We apply convergent iterative k-core filtering, repeatedly removing users with fewer than Minimum User Interactions (MUI) and items with fewer than Min Item Interactions (MII) until no further removals occur. This procedure guarantees that all remaining users and items simultaneously satisfy the specified interaction thresholds. We evaluate five k-core configurations, denoted as $(u,i)$, where $u$ and $i$ represent the minimum number of interactions required per user and per item, respectively, as shown in Table~\ref{tab:kcore_filtering}.

Our primary benchmarking configuration is $(10,3)$, which retains 876,433 interactions. This setting provides a favorable trade-off between dataset size and interaction density while preserving sufficient observations for both user-based and item-based collaborative filtering methods. The resulting dataset contains 39,550 users and 31,713 items, yielding a relatively balanced user--item ratio.

\begin{table}
\centering
\caption{K-core filtering configurations and resulting dataset sparsity. MUI and MII indicates minimum interaction per user and item respectively.}
\label{tab:kcore_filtering}
\small
\begin{tabular}{cccccc}
\toprule
(MUI, MII) & Users & Items & Interactions & Density & Retention \\
\midrule
(5, 3)  & 107,075 & 42,462 & 1,327,720 & 0.029\% & 49.7\% \\
(10, 3) & 39,550  & 31,713 & 876,433   & 0.069\% & 32.8\% \\
(15, 3) & 20,957  & 25,074 & 647,838   & 0.12\% & 24.3\% \\
(15, 5) & 19,032  & 15,022 & 587,828   & 0.19\% & 22.0\% \\
(20, 5) & 11,454  & 11,499 & 447,773   & 0.32\% & 16.8\% \\
\bottomrule
\end{tabular}
\end{table}

Each successive threshold substantially reduces coverage — the primary (10,3) configuration retains 32.8\% of interactions and 4.2\% of the original user population, concentrating evaluation on the platform's most active users and representing a practical limitation on result generalizability.

\textbf{Temporal Split:}
For each user, interactions are sorted chronologically by timestamp. The most recent interaction is assigned to the test set, the second-most recent interaction to the validation set, and all preceding interactions to the training set. This leave-last-out protocol simulates next-interaction prediction while preventing future interactions from contaminating model training \cite{meng2020exploringdatasplittingstrategies, Gusak_2025}. Explicit leakage checks verify zero overlap among training, validation, and test partitions. After k-core filtering, users with fewer than three interactions are excluded because at least one interaction is required for each of the training, validation, and test splits.

%% file: Sources/experiments.tex
\section{Experimental Setup}
\subsection{Models}

\textbf{GlobalPop:} A non-personalized baseline that recommends the globally most-interacted items to all users \cite{10.1145/3397271.3401233}.

\textbf{CatPop:} A category-aware popularity baseline that identifies each user's dominant training category and recommends the most popular items within that category. For users with insufficient interaction history, recommendations default to GlobalPop \cite{10.1007/s10618-022-00913-5}.

\textbf{UserCF:} A neighborhood-based collaborative filtering model that constructs a sparse user--item interaction matrix, computes pairwise cosine similarity between users, and aggregates preferences from the top-50 most similar neighbors to generate recommendations \cite{10.1145/371920.372071}.

\textbf{ItemCF:} An item-based collaborative filtering model that computes cosine similarity between items and scores candidate items using similarity-weighted aggregation over a user's interaction history \cite{10.1145/371920.372071}. ItemCF is expected to be more robust than UserCF under extreme sparsity because item similarities are generally more stable than user similarities.

\textbf{ExplicitMF (SVD):} A matrix factorization model that decomposes the observed rating matrix into latent user and item representations by minimizing squared reconstruction error \cite{10.1109/MC.2009.263}. We use 100 latent factors and an $\ell_2$ regularization coefficient of 0.02.

\textbf{ImplicitMF (ALS):} An implicit-feedback matrix factorization model that treats interactions as binary preference signals with confidence weighting and optimizes latent representations using Alternating Least Squares (ALS) \cite{10.1109/ICDM.2008.22}. The model uses 64 latent factors.

\begin{table*}
\centering
\caption{Model performance at the primary (10,3) setting (39,550 users, 31,713 items). Best results are bolded.}
\label{tab:model_performance_10_3}
\small
\begin{tabular}{lccccccccc}
\toprule
Model & Hit@10 & Hit@30 & Hit@50 & MRR@10 & MRR@30 & MRR@50 & NDCG@10 & NDCG@30 & NDCG@50 \\
\midrule
GlobalPop & 0.0215 & 0.046 & 0.0613 & 0.006 & 0.0072 & 0.0076 & 0.0096 & 0.0151 & 0.0179 \\
CatPop & 0.0202 & 0.037 & 0.0468 & 0.0076 & 0.0085 & 0.0087 & 0.0105 & 0.0145 & 0.0163 \\
UserCF & 0.0444 & 0.081 & 0.108 & 0.0167 & 0.0187 & 0.0194 & 0.0231 & 0.0317 & 0.0368 \\
\textbf{ItemCF} & \textbf{ 0.0525} & \textbf{ 0.095} & \textbf{0.1241} & \textbf{0.0194} & \textbf{0.0219} & \textbf{0.0226} & \textbf{0.0270} & \textbf{0.0373} & \textbf{0.0426} \\
ExplicitMF & 0.0361 & 0.066 & 0.0877 & 0.0148 & 0.0165 & 0.017 & 0.0198 & 0.0269 & 0.0309 \\
ImplicitMF & 0.0399 & 0.082 & 0.1123 & 0.0135 & 0.0159 & 0.0167 & 0.0196 & 0.0296 & 0.0352 \\
\bottomrule
\end{tabular}
\end{table*}

\textbf{Content-Based Filtering (CBF):} A content-based recommender that represents items using TF--IDF features extracted from aggregated review text. To improve robustness to Banglish transliteration variation, we employ character 1--2 gram features with a vocabulary size of 15{,}000 \cite{10129187}. When review text is unavailable, a three-stage fallback strategy is applied: (1) TF--IDF features derived from review text, (2) category-string features, and (3) a zero vector representation. The final fallback applies to approximately 8\% of items. User profiles are constructed as the $\ell_2$-normalized mean of the vectors corresponding to items in the user's training history \cite{10.5555/1768197.1768209}. Due to computational constraints, CBF is evaluated only under the primary (10,3) filtering configuration.

\subsection{Evaluation Metrics}

Recommendation quality is evaluated using Hit Rate (HR@K), Mean Reciprocal Rank (MRR@K), and Normalized Discounted Cumulative Gain (NDCG@K) at $K \in \{10, 30, 50\}$. Among these metrics, NDCG@10 serves as the primary evaluation criterion. To ensure fair evaluation, items observed during training are excluded from each user's candidate set at inference time. All metrics are computed using a vectorized NumPy-based evaluation pipeline.

For the code-mixing analysis, users are partitioned according to the dominant script used in their review history (e.g., Bangla, English, or Banglish). NDCG@10 is then computed separately for each user segment to assess whether recommendation performance varies across different linguistic usage patterns.

\section{Results and Analysis}
\subsection{Primary Benchmark (Min User/Item = 10,3)}

ItemCF achieves the best overall performance across all evaluation metrics and cutoff values, attaining an NDCG@10 of 0.027 and an NDCG@50 of 0.043. UserCF ranks second at every cutoff, establishing neighborhood-based collaborative filtering as the strongest model family on this benchmark. This result is consistent with prior observations that latent-factor methods often struggle in extremely sparse settings, where limited interaction data hinders the estimation of reliable user and item representations \cite{10.1109/ICDM.2008.22}.

The two matrix factorization approaches exhibit different ranking characteristics. ExplicitMF slightly outperforms ImplicitMF at short recommendation lists (NDCG@10: 0.0198 vs.\ 0.0196), whereas ImplicitMF surpasses ExplicitMF at larger cutoffs (NDCG@30: 0.0296 vs.\ 0.0269), with the performance gap increasing further at K=50. This pattern suggests that ExplicitMF produces more accurate top-ranked recommendations, while ImplicitMF achieves greater coverage and recall over longer recommendation lists.

Finally, the non-personalized baselines perform substantially worse than all personalized methods across every cutoff. Despite the extreme sparsity of the interaction matrix, user-specific interaction histories still provide sufficient signal to support meaningful personalization.

\subsection{Sparsity Ablation}

\begin{table}[t]
\centering
\caption{Ablation for NDCG@10 across k-core threshold configurations.}
\label{tab:ndcg_kcore}
\small
\begin{tabular}{lcccccc}
\toprule
Model & (5,3) & (10,3) & (15,3) & (15,5) & (20,5) & $\Delta$ \\
\midrule
GlobalPop & 0.0078 & 0.0096 & 0.0106 & 0.0118 & 0.0123 & +0.0045 \\
CatPop & 0.0126 & 0.0105 & 0.0102 & 0.0114 & 0.0122 & $-$0.0004 \\
UserCF & 0.0305 & 0.0231 & 0.0199 & 0.0215 & 0.0191 & $-$0.0114 \\
ItemCF & 0.0302 & 0.0270 & 0.0244 & 0.0270 & 0.0255 & $-$0.0047 \\
ExplicitMF & 0.0200 & 0.0198 & 0.0201 & 0.0228 & 0.0243 & +0.0043 \\
ImplicitMF & 0.0275 & 0.0196 & 0.0156 & 0.0170 & 0.0136 & $-$0.0138 \\
\bottomrule
\end{tabular}
\end{table}

The density ablation reveals four distinct sensitivity profiles, indicating that recommendation algorithms respond differently to increasingly sparse interaction data. \textbf{ItemCF} is the most robust model, exhibiting only a modest performance decline ($-0.0047$). This stability is consistent with item-based similarity estimation being less sensitive to reductions in user coverage than user-based approaches.

In contrast, \textbf{ImplicitMF} is the most adversely affected ($-0.0138$). A plausible explanation is that confidence-weighted implicit-feedback models require sufficient interaction volume to differentiate user preferences; under extreme sparsity, confidence values become increasingly homogeneous, reducing the model's ability to learn discriminative latent representations.

\textbf{ExplicitMF} exhibits the opposite trend, improving as density increases ($+0.0043$). One possible explanation is the highly skewed rating distribution, where 77.93\% of ratings are five stars. Under sparse conditions, this skew may encourage near-constant rating predictions. As denser user--item interactions are retained, greater rating diversity becomes available, yielding a more informative supervision signal for factorization.

 \textbf{UserCF} experiences a moderate performance degradation ($-0.0114$), consistent with the well-documented sensitivity of user-based neighborhood methods to reductions in user overlap. Finally, \textbf{GlobalPop} benefits from increasing density, suggesting that popularity estimates become more reliable as interaction counts accumulate.

\subsection{Code-Mix Impact on Content-Based Filtering}
\begin{table}[t]
\centering
\caption{CBF performance by user dominant script (10,3 setting, 39,550 users).}
\label{tab:cbf_script_performance}
\small
\begin{tabular}{lrrrr}
\toprule
Script Segment & N Users & NDCG@10 & Hit@10 & $\Delta$ vs Bengali \\
\midrule
Bengali & 5,354 & 0.0094 & 0.0211 & -- \\
Empty (no reviews) & 24,183 & 0.0081 & 0.0166 & $-$13.8\% \\
English & 9,096 & 0.0062 & 0.0133 & $-$34.0\% \\
Banglish & 866 & 0.0050 & 0.0115 & $-$46.8\% \\
\textbf{Overall CBF} & \textbf{39,550} & \textbf{0.0077} & \textbf{0.0160} & -- \\
\bottomrule
\end{tabular}
\end{table}

At the (10,3) threshold, ItemCF achieves NDCG@10 = 0.027 while CBF achieves 0.0077 —
a 3.5× gap demonstrating that collaborative signals substantially outperform text-based
profiles in this domain.

Within CBF, Banglish-dominant users suffer the steepest degradation ($-46.8$\% relative to
Bengali). This is directly attributable to transliteration inconsistency: a single concept
is split across multiple surface forms (e.g., *valo/bhalo/balo/vhalo* for "good";
*sundor/shundor* for "nice"; *khub/kub/khob* for "very"), fragmenting the
TF-IDF vocabulary and reducing inter-item similarity. Character 1-2 gram features
partially mitigate this, but cannot fully bridge orthographic distance across arbitrary
romanization choices. English-dominant users also degrade meaningfully ($-34.0$\%), likely
due to shorter average review length (mean 40.4 chars vs 73.9 for Bengali) and higher
generic vocabulary overlap reducing discrimination.

A counterintuitive finding: users with no dominant review script ("empty") outperform
both English and Banglish users (NDCG@10 = 0.0081). Since user profiles are computed as
the mean of their interacted item vectors, "empty" users benefit from item profiles
constructed from *other users'* text — circumventing the transliteration fragmentation
problem by never relying on their own noisy textual history.

%% file: Sources/related.tex
\section{Related Works}

\begin{table}
\centering
\caption{Comparison of recommendation benchmarks in the Bangla and global domain.}
\label{tab:benchmark_comparison}
\small
\begin{tabular}{llllll}
\toprule
Work & Domain & Language & Interactions & Code-Mix \\
\midrule
MovieLens \cite{10.1145/2827872} & Movies & English & 20M & $\times$ \\
Amazon \cite{10.1145/2872427.2883037} & Fashion & English & $\sim$1M & $\times$ \\
RokomariBG \cite{ahmed2026personalizedbanglabookrecommendation} & Books & Bangla & 210K & $\times$ \\
BLaIR \cite{hou2026bridginglanguageitemsretrieval} & E-comm & English & 570M & $\times$ \\
\textbf{This work} & \textbf{E-comm} & \textbf{BN+EN+Mix} & \textbf{2.67M} & \textbf{$\checkmark$} \\
\bottomrule
\end{tabular}
\end{table}

Recommendation benchmarking has been anchored to a small set of English-language
datasets. MovieLens \cite{10.1145/2827872} and Amazon Reviews \cite{10.1145/2872427.2883037} remain dominant testbeds for
collaborative filtering \cite{10.1145/371920.372071, 10.1109/ICDM.2008.22} and matrix factorization \cite{10.1109/MC.2009.263, 10.5555/1795114.1795167} research. In the
Bangla domain, RokomariBG \cite{ahmed2026personalizedbanglabookrecommendation} is the first published recommendation benchmark, evaluating
models from CF and MF to LightGCN and neural two-tower retrieval on Bangla book data.
However, it does not study code-mixed text
effects. Prior Bangladeshi e-commerce datasets are predominantly constructed for
sentiment analysis \cite{10456378, RASHID2024110052, 9392733}, lacking the user-item interaction structure
required for personalized recommendation. BLaIR \cite{hou2026bridginglanguageitemsretrieval} benchmarks LLMs as semantic encoders
across 570M English reviews, finding poor correlation with general embedding benchmarks —
reinforcing that recommendation is a distinct task requiring domain-specific evaluation.
For code-mixed NLP, shared tasks on Dravidian languages \cite{10.1145/3503162.3503177} and Bangla-English mixing \cite{10129187}
establish that transliteration inconsistency severely fragments vocabulary in word-level
models, motivating character n-gram approaches \cite{ISLAM2024100069}.

%% file: Sources/conclusion.tex
\section{Conclusion}

This work presents the first Top-$N$ recommendation benchmarking study on interaction data collected from Daraz Bangladesh, the country's largest e-commerce platform. As Bangladesh's e-commerce sector continues to expand, effective recommender systems are becoming increasingly important for product discovery and customer engagement. Our benchmarking results provide an empirical foundation for understanding recommendation performance in this emerging market setting.

Across five $k$-core density configurations and a temporally consistent evaluation protocol, neighborhood-based collaborative filtering methods consistently outperform matrix factorization approaches, highlighting the challenges posed by extreme interaction sparsity. The density ablation further provides practical deployment insights: ItemCF is the most robust under sparse conditions, ExplicitMF benefits from increased interaction density, and ImplicitMF is particularly sensitive to sparsity.

Beyond benchmarking, our language-stratified analysis identifies Banglish vocabulary fragmentation as a challenge for content-based recommendation. The coexistence of Bangla, English, and multiple Banglish transliterations introduces linguistic variability that degrades text-based item representations. This finding motivates future work on transliteration normalization, multilingual representations, and language-aware recommender systems for Bengali and other low-resource multilingual markets.

%% file: references.bib
@misc{shamael2024banglishrevlargescalebanglaenglishcodemixed,
      title={BanglishRev: A Large-Scale Bangla-English and Code-mixed Dataset of Product Reviews in E-Commerce}, 
      author={Mohammad Nazmush Shamael and Sabila Nawshin and Swakkhar Shatabda and Salekul Islam},
      year={2024},
      eprint={2412.13161},
      archivePrefix={arXiv},
      primaryClass={cs.CL},
      url={https://arxiv.org/abs/2412.13161}, 
}

@misc{bangladeshi-ecommerce,
  author = {GLOBE NEWSWIRE},
  title = {Bangladesh B2C Ecommerce Market Forecast Report 2025-2029},
  year = {2026},
  url = {https://www.globenewswire.com/news-release/2026/01/29/3228360/28124/en/bangladesh-b2c-ecommerce-market-forecast-report-2025-2029-daraz-leads-reach-while-chaldal-and-pickaboo-anchor-key-verticals-as-new-social-commerce-and-b2b2c-entrants-intensify-comp.html},
  note = {Accessed: 2026-06-01}
}

@incollection{ricci2010introduction,
  author       = {Francesco Ricci and
                  Lior Rokach and
                  Bracha Shapira},
  editor       = {Francesco Ricci and
                  Lior Rokach and
                  Bracha Shapira and
                  Paul B. Kantor},
  title        = {Introduction to Recommender Systems Handbook},
  booktitle    = {Recommender Systems Handbook},
  pages        = {1--35},
  publisher    = {Springer},
  address      = {Boston, MA, USA},
  year         = {2011},
  url          = {https://doi.org/10.1007/978-0-387-85820-3\_1},
  doi          = {10.1007/978-0-387-85820-3\_1},
  bibsource    = {dblp computer science bibliography, https://dblp.org}
}

@misc{ahmed2026personalizedbanglabookrecommendation,
      title={Towards Personalized Bangla Book Recommendation: A Large-Scale Multi-Entity Book Graph Dataset}, 
      author={Rahin Arefin Ahmed and Md. Anik Chowdhury and Sakil Ahmed Sheikh Reza and Devnil Bhattacharjee and Muhammad Abdullah Adnan and Nafis Sadeq},
      year={2026},
      eprint={2602.12129},
      archivePrefix={arXiv},
      primaryClass={cs.IR},
      url={https://arxiv.org/abs/2602.12129}, 
}

@inproceedings{10.1145/3503162.3503177,
author = {Priyadharshini, Ruba and Chakravarthi, Bharathi raja and Thavareesan, Sajeetha and Chinnappa, Dhivya and Thenmozhi, Durairaj and Ponnusamy, Rahul},
title = {Overview of the DravidianCodeMix 2021 Shared Task on Sentiment Detection in Tamil, Malayalam, and Kannada},
year = {2022},
isbn = {9781450395960},
publisher = {Association for Computing Machinery},
address = {New York, NY, USA},
url = {https://doi.org/10.1145/3503162.3503177},
doi = {10.1145/3503162.3503177},
booktitle = {Proceedings of the 13th Annual Meeting of the Forum for Information Retrieval Evaluation},
pages = {4–6},
numpages = {3},
location = {Virtual Event, India},
series = {FIRE '21}
}

@article{RASHID2024110052,
title = {A comprehensive dataset for sentiment and emotion classification from Bangladesh e-commerce reviews},
journal = {Data in Brief},
volume = {53},
pages = {110052},
year = {2024},
issn = {2352-3409},
doi = {https://doi.org/10.1016/j.dib.2024.110052},
url = {https://www.sciencedirect.com/science/article/pii/S235234092400026X},
author = {Mohammad Rifat Ahmmad Rashid and Kazi Ferdous Hasan and Rakibul Hasan and Aritra Das and Mithila Sultana and Mahamudul Hasan}
}

@INPROCEEDINGS{10456378,
  author={Mukit, Mohammad and Rabbi, Md Moshiul Huq and Ahmed, Mohammad Masud and Latif, Subhenur and Saha, Sajib},
  booktitle={2023 IEEE 9th International Women in Engineering (WIE) Conference on Electrical and Computer Engineering (WIECON-ECE)}, 
  title={Sentiment Analysis on Bangla and Phonetic Bangla Reviews: A Product Rating Procedure using NLP and Machine Learning}, 
  year={2023},
  volume={},
  number={},
  pages={433-437},
  doi={10.1109/WIECON-ECE60392.2023.10456378}}

@ARTICLE{10129187,
  author={Tareq, Mohammad and Islam, Md. Fokhrul and Deb, Swakshar and Rahman, Sejuti and Mahmud, Abdullah Al},
  journal={IEEE Access}, 
  title={Data-Augmentation for Bangla-English Code-Mixed Sentiment Analysis: Enhancing Cross Linguistic Contextual Understanding}, 
  year={2023},
  volume={11},
  number={},
  pages={51657-51671},
  doi={10.1109/ACCESS.2023.3277787}}

@misc{hou2026bridginglanguageitemsretrieval,
      title={Bridging Language and Items for Retrieval and Recommendation: Benchmarking LLMs as Semantic Encoders}, 
      author={Yupeng Hou and Jiacheng Li and Xiangjun Fu and Zhankui He and An Yan and Xiusi Chen and Julian McAuley},
      year={2026},
      eprint={2403.03952},
      archivePrefix={arXiv},
      primaryClass={cs.IR},
      url={https://arxiv.org/abs/2403.03952}, 
}

@inproceedings{10.1145/371920.372071,
author = {Sarwar, Badrul and Karypis, George and Konstan, Joseph and Riedl, John},
title = {Item-based collaborative filtering recommendation algorithms},
year = {2001},
isbn = {1581133480},
publisher = {Association for Computing Machinery},
address = {New York, NY, USA},
url = {https://doi.org/10.1145/371920.372071},
doi = {10.1145/371920.372071},
booktitle = {Proceedings of the 10th International Conference on World Wide Web},
pages = {285–295},
numpages = {11},
location = {Hong Kong, Hong Kong},
series = {WWW '01}
}

@inproceedings{10.1109/ICDM.2008.22,
author = {Hu, Yifan and Koren, Yehuda and Volinsky, Chris},
title = {Collaborative Filtering for Implicit Feedback Datasets},
year = {2008},
isbn = {9780769535029},
publisher = {IEEE Computer Society},
address = {USA},
url = {https://doi.org/10.1109/ICDM.2008.22},
doi = {10.1109/ICDM.2008.22},
booktitle = {Proceedings of the 2008 Eighth IEEE International Conference on Data Mining},
pages = {263–272},
numpages = {10},
series = {ICDM '08}
}

@article{10.1109/MC.2009.263,
author = {Koren, Yehuda and Bell, Robert and Volinsky, Chris},
title = {Matrix Factorization Techniques for Recommender Systems},
year = {2009},
issue_date = {August 2009},
publisher = {IEEE Computer Society Press},
address = {Washington, DC, USA},
volume = {42},
number = {8},
issn = {0018-9162},
url = {https://doi.org/10.1109/MC.2009.263},
doi = {10.1109/MC.2009.263},
journal = {Computer},
month = aug,
pages = {30–37},
numpages = {8}
}

@inproceedings{10.5555/1795114.1795167,
author = {Rendle, Steffen and Freudenthaler, Christoph and Gantner, Zeno and Schmidt-Thieme, Lars},
title = {BPR: Bayesian personalized ranking from implicit feedback},
year = {2009},
isbn = {9780974903958},
publisher = {AUAI Press},
address = {Arlington, Virginia, USA},
booktitle = {Proceedings of the Twenty-Fifth Conference on Uncertainty in Artificial Intelligence},
pages = {452–461},
numpages = {10},
location = {Montreal, Quebec, Canada},
series = {UAI '09}
}

@misc{meng2020exploringdatasplittingstrategies,
      title={Exploring Data Splitting Strategies for the Evaluation of Recommendation Models}, 
      author={Zaiqiao Meng and Richard McCreadie and Craig Macdonald and Iadh Ounis},
      year={2020},
      eprint={2007.13237},
      archivePrefix={arXiv},
      primaryClass={cs.IR},
      url={https://arxiv.org/abs/2007.13237}, 
}

@inproceedings{Gusak_2025, series={RecSys ’25},
   title={Time to Split: Exploring Data Splitting Strategies for Offline Evaluation of Sequential Recommenders},
   url={http://dx.doi.org/10.1145/3705328.3748164},
   DOI={10.1145/3705328.3748164},
   booktitle={Proceedings of the Nineteenth ACM Conference on Recommender Systems},
   publisher={ACM},
   author={Gusak, Danil and Volodkevich, Anna and Klenitskiy, Anton and Vasilev, Alexey and Frolov, Evgeny},
   year={2025},
   month=Sept, pages={874–883},
   collection={RecSys ’25} }

@inbook{10.5555/1768197.1768209,
author = {Pazzani, Michael J. and Billsus, Daniel},
title = {Content-based recommendation systems},
year = {2007},
isbn = {9783540720782},
publisher = {Springer-Verlag},
address = {Berlin, Heidelberg},
booktitle = {The Adaptive Web: Methods and Strategies of Web Personalization},
pages = {325–341},
numpages = {17}
}

@inproceedings{10.1145/2872427.2883037,
author = {He, Ruining and McAuley, Julian},
title = {Ups and Downs: Modeling the Visual Evolution of Fashion Trends with One-Class Collaborative Filtering},
year = {2016},
isbn = {9781450341431},
publisher = {International World Wide Web Conferences Steering Committee},
address = {Republic and Canton of Geneva, CHE},
url = {https://doi.org/10.1145/2872427.2883037},
doi = {10.1145/2872427.2883037},
booktitle = {Proceedings of the 25th International Conference on World Wide Web},
pages = {507–517},
numpages = {11},
location = {Montr\'{e}al, Qu\'{e}bec, Canada},
series = {WWW '16}
}

@INPROCEEDINGS{9392733,
  author={Shafin, Minhajul Abedin and Hasan, Md. Mehedi and Alam, Md. Rejaul and Mithu, Mosaddek Ali and Nur, Arafat Ulllah and Faruk, Md. Omar},
  booktitle={2020 23rd International Conference on Computer and Information Technology (ICCIT)}, 
  title={Product Review Sentiment Analysis by Using NLP and Machine Learning in Bangla Language}, 
  year={2020},
  volume={},
  number={},
  pages={1-5},
  doi={10.1109/ICCIT51783.2020.9392733}
  }

@INPROCEEDINGS{9033741,
  author={Sarowar, Md. Golam and Rahman, Mushfiqur and Yousuf Ali, Md. Nawab and Rakib, Omor Faruk},
  booktitle={2019 IEEE 5th International Conference for Convergence in Technology (I2CT)}, 
  title={An Automated Machine Learning Approach for Sentiment Classification of Bengali E-Commerce Sites}, 
  year={2019},
  volume={},
  number={},
  pages={1-5},
  doi={10.1109/I2CT45611.2019.9033741}
  }

@article{ISLAM2024100069,
title = {Sentiment analysis of Bangla language using a new comprehensive dataset BangDSA and the novel feature metric skipBangla-BERT},
journal = {Natural Language Processing Journal},
volume = {7},
pages = {100069},
year = {2024},
issn = {2949-7191},
doi = {https://doi.org/10.1016/j.nlp.2024.100069},
url = {https://www.sciencedirect.com/science/article/pii/S2949719124000177},
author = {Md. Shymon Islam and Kazi Masudul Alam}
}

@article{10.1145/2827872,
author = {Harper, F. Maxwell and Konstan, Joseph A.},
title = {The MovieLens Datasets: History and Context},
year = {2015},
issue_date = {January 2016},
publisher = {Association for Computing Machinery},
address = {New York, NY, USA},
volume = {5},
number = {4},
issn = {2160-6455},
url = {https://doi.org/10.1145/2827872},
doi = {10.1145/2827872},
journal = {ACM Trans. Interact. Intell. Syst.},
month = dec,
articleno = {19},
numpages = {19}
}

@inproceedings{10.1145/3397271.3401233,
author = {Ji, Yitong and Sun, Aixin and Zhang, Jie and Li, Chenliang},
title = {A Re-visit of the Popularity Baseline in Recommender Systems},
year = {2020},
isbn = {9781450380164},
publisher = {Association for Computing Machinery},
address = {New York, NY, USA},
url = {https://doi.org/10.1145/3397271.3401233},
doi = {10.1145/3397271.3401233},
booktitle = {Proceedings of the 43rd International ACM SIGIR Conference on Research and Development in Information Retrieval},
pages = {1749–1752},
numpages = {4},
location = {Virtual Event, China},
series = {SIGIR '20}
}

@article{10.1007/s10618-022-00913-5,
author = {S\'{a}nchez, Pablo and Bellog\'{\i}n, Alejandro and Boratto, Ludovico},
title = {Bias characterization, assessment, and mitigation in location-based recommender systems},
year = {2023},
issue_date = {Sep 2023},
publisher = {Kluwer Academic Publishers},
address = {USA},
volume = {37},
number = {5},
issn = {1384-5810},
url = {https://doi.org/10.1007/s10618-022-00913-5},
doi = {10.1007/s10618-022-00913-5},
journal = {Data Min. Knowl. Discov.},
month = feb,
pages = {1885–1929},
numpages = {45}
}

@techreport{united_nations_world_2024,
	address = {New York},
	type = {Technical {Report}},
	title = {World {Population} {Prospects} 2024: {Summary} of {Results}},
	url = {https://desapublications.un.org/publications/world-population-prospects-2024-summary-results},
	number = {UN DESA/POP/2024/TR/NO. 9},
	institution = {United Nations, Department of Economic and Social Affairs, Population Division},
	author = {{United Nations}},
	year = {2024},
}


%% file: sample-base.bib
@String{Computing = "Computing" }

@String{Computer = "{IEEE} Computer" }

@String{Academic = "Academic Press" }

@String{Springer = "Springer-Verlag" }
